\documentstyle[12pt,psfig,aaspp4]{article}
\begin{document}

\title{Astrometric Shifts in the OGLE-1 Microlensing Events}

\author{David M. Goldberg and Przemys\l aw R. Wo\'zniak}
\affil{Princeton University Observatory, Princeton, NJ 08544-1001}
\affil{email: (goldberg, wozniak)@astro.princeton.edu}

\begin{abstract}
We measure the astrometric shifts of the light centroids for the microlensing
events in the OGLE-1 database.  Of the $15$ consistently detected events, $7$
had significant shifts which we were able to measure with a fair degree of
confidence.  Those events with large shifts are also expected to be unresolved
with a background ``blend'', and thus, we suggest that we have identified
events in the OGLE-1 catalog which are strongly blended.  Though we concentrate
on the OGLE-1 database, and use the DoPHOT package in order to perform our
analysis, we suggest that this shift is a generic effect, and should be
observable in any crowded field.
\end{abstract}


\section{Introduction}

The search for gravitational microlensing events has been very successful. To
date, over $100$ candidate events have been detected, primarily by the MACHO
collaboration (Alcock et al. 1993), but also significantly by the OGLE (Udalski
et al. 1992), DUO (Alard 1996a), and EROS (Palanque-Delabrouille et al. 1997)
collaborations.  In order to maximize event rates, the searches take place
toward crowded fields such as the Galactic bulge and the LMC (Alcock et
al. 1996).

While searches toward crowded fields allow for a large number of detections,
this also makes it highly likely that an observed star will have several
contributing unresolved sources.  Since only one of these sources is lensed at
any given time, events will, in general, be blended (Di Stefano \& Esin 1995;
Alard 1997; Wo\'{z}niak \& Paczy\'{n}ski 1997; Han 1997; Goldberg 1998).

For any event, we define the blending fraction, $f$ as:
\begin{equation}
f\equiv \frac{F_{s}}{\sum_{i}F_{i}}=\frac{F_{s}}{F_{0}} \leq 1 \ ,
\end{equation}
where $F_{s}$ is the flux from the lensed star, $F_{i}$ is the flux from each
of the contributing stars in the point spread function (PSF), and $F_{0}$ is
the total flux from the observed object.

If an event is blended, a measurement of the microlensing magnification,
$A_{obs}(t)$, will be a systematic underestimate of the true magnification of
the source star, $A(t)$:
\begin{equation}
A_{obs}(t)=1-f+A(t)f \leq A(t)\ .
\end{equation}

For a single lens microlensing event, $A(t)$ evolves in a straightforward way
(Paczy\'nski 1986; 1996):
\begin{eqnarray}
\label{eq:lcurve}
A(t)=\frac{u(t)^{2}+2}{u(t)\sqrt{u(t)^{2}+4}} & ; & u(t)^{2}=u_{min}^{2}+
\left( \frac{t-t_{max}}{t_{0}} \right)^{2}\ ,
\end{eqnarray}

where $u_{min}$ is the impact parameter of the lensed star with respect to the
lens in units of the Einstein radius, $t_{0}$ is the characteristic time of
the event, and $t_{max}$ is the time of maximum magnification.  Wo\'zniak \&
Paczy\'nski (1997) show that the parameters \{ $f$, $u_{min}$, $t_{0}$ \}
form an approximately degenerate set, and since $f$ is not known independently,
the other parameters cannot be known either.

It has been suggested by one of us (Goldberg 1998) that in very crowded fields
a large fraction of gravitational microlensing events will exhibit a
characteristic astrometric shift. If an observed image is composed of two or
more true stars, then the center of the PSF will be weighted toward the center
of light of contributing stars.  For an unblended or very weakly blended event
($f\simeq 1$), the center of light is essentially defined by the position of
the lensed star.  However, for a strongly blended event ($f\ll 1$) the lensed
star may be almost anywhere within the PSF and not significantly affect the
center of light.

As the magnification increases, the center of light will shift toward the
lensed star. Thus, if we consider a blended image which is centered at the
origin at baseline flux, and then lens a star within the image at $(\Delta
x_{0},\Delta y_{0})$, simple geometrical arguments show that the center of
light will shift toward the lensed star:

\begin{eqnarray}
\Delta x=\Delta x_{0}\frac{A_{obs}(t)-1}{A_{obs}(t)} & ; &
\Delta y=\Delta y_{0}\frac{A_{obs}(t)-1}{A_{obs}(t)}\ .
\label{eq:lines}
\end{eqnarray}

This effect has been observed; Alard (1996b) demonstrated a shift in DU0 \#
2, a binary event which was known to be blended photometrically.  However,
astrometric shifts have not been more widely used, and indeed, there is a
certain amount of skepticism that they may be significant (Han 1997).

Goldberg (1998) suggested that by looking at astrometric shift information from
a microlensing event, we may approach the problem of blending in two ways.
First, those events with larger shifts will tend to have a higher degree of
blending than those with small shifts. Detailed models of a population can
potentially yield a loose shift-blending relation. Second, by performing a
best fit of equations~(\ref{eq:lines}) for $\Delta x_{0}$ and $\Delta y_{0}$,
we can find the position of the true lensed star with respect to the observed
PSF and the local field. High resolution followup observations can then be
used to determine the brightness of the lensed star, and give a direct measure
of the degree of blending.

In this paper, we shall measure the astrometric shift for candidate
microlensing events in the OGLE archive.  Though we do not attempt to
explicitly to calculate it here, any quantitative estimate of the optical depth
of the galaxy to microlensing must include the effect of blending, and the
presence or absence of a shift can provide a handle on this.  Our outline is as
follows. In \S~\ref{sec:analysis}, we discuss our analysis methods, in
particular, data reduction, photometry, and determination of the position
shift.  In \S~\ref{sec:results} we discuss the astrometric shifts found for the
OGLE archive. Finally, in \S~\ref{sec:discussion}, we discuss some interesting
effects that we noted in this study as well as present some suggestions for
future study.

\section{Analysis}
\label{sec:analysis}

\subsection{Data}
\label{sec:data}

Events analyzed in this paper are those reported by OGLE collaboration from the
first stage of the experiment (Udalski et al. 1992) including objects caught on
the spot by the ``early warning system'' (Udalski et al. 1994b).  Wo\'{z}niak
and Szyma\'{n}ski (1997; hereafter W\& S) extracted from every frame in OGLE-1
archive a $100 \times 100$ pix subframe centered on objects which were formally
variable only in one observing season and constant during remaining
seasons. This dataset includes OGLE-1 microlensing candidates and is restricted
to good quality frames (grades A-E, F frames rejected).  Depending on the field
it covers 3 or 4 seasons between 1992 and 1995, of approximately 4 months of
observing each. The single Ford (Loral) $2048 \times 2048$ CCD with $15 \mu m$
pixels used by OGLE team results in a scale of 0.44 arcsec/pix.  A $44 \times
44$ arcsec subframe of a dense stellar field towards the Galactic center
contains plenty of stars to allow difference photometry and good registering of
the images.  The images, as provided by W\& S were de-biased and flat-fielded
by an automated data pipe line based on IRAF "ccdred" routines. We refer to
Udalski et al. (1992) and W\&S for details of preliminary reductions, selection
criteria and occasional exceptions.

\subsection{Method}

All of the remaining steps, that is profile photometry, cross identification of
stars and registering of images were done independently.  Previous data
analysis of microlensing events has focused on identifying candidate events and
using the best photometry possible to determine a light curve (and hence the
microlensing parameters).  The OGLE team used a fixed centroid template to
``warm start'' DoPHOT (Schechter et al. 1995) in order to maximize
identification rate of the stars in dense fields (equivalent to the assumption
that there is no shift !).  To find the shift, however, we need to define an
accurate ``absolute'' reference coordinate system and measure the position of
the amplified star with respect to this system. Therefore we use DoPHOT in
different mode and identify stars in each frame independently.

The outline of our method is as follows.
For each of the $100 \times 100$ pix image, we create an observed catalog,
that is the list of stars found by DoPHOT.
The brightnesses and positions of stars we found in all subframes during
unlensed period are used to generate a template catalog, a basis of the
reference coordinate system. During the lensed period, the positions
of the lensed star relative to the template are computed.
Finally, errors in position and brightness were determined, and a best fit of
equations~(\ref{eq:lcurve}) and (\ref{eq:lines}) were computed.

\subsection{Locating Stars and Photometry}

\label{sec:method}

We benefit from the analysis by the OGLE group in many ways, not in
the least because we did not need to calculate a seeing and approximate
sky level for each subframe, as this had already been done.

We divide the subframes into two categories: those during the event, and those
preceding or following the event.  Again, we have the benefit of the previously
computed light curves.  For those outside the event period
(typically $|t-t_{max}|\gtrsim 2\ t_{0}$), we select one of the ``A'' quality
frames to serve as an initial template. Note that this frame is privileged
primarily in the sense that all other frames will be rotated into the same
coordinates.  

The template catalog is adjusted iteratively.  We begin with one frame, and add
additional frames by determining a four parameter coordinate transform: $\{
\theta_{i}, \delta x_{i}, \delta y_{i}, F_{i}/F_{1} \}$, that is
rotation, parallel shift and flux ratio.

We determine the values of the transformation vectors by comparing the new
stars to the template in order of decreasing brightness.  As each star is
added, it is rotated using the transformation matrix and compared to the
previous template, and if the fit is acceptably good (typically $\Delta r<2$
pixels and differences in flux less than $20\%$), we iteratively adjust the
transformation vector to minimize $\chi^{2}$ over all the new stars.
Otherwise, we throw out the star.  This continues through the entire catalog.
Typically, we were able to associate $\sim 120$ stars per subframe with the
stars in the template.  Once we have found the transformation matrix and
assigned the new stars to the template stars, we recompute the position and
brightness of each template star by taking the reference system resulting from
an average of all the subframes observed thus far.

We perform this comparison for all the baseline subframes (typically $\sim
100$), and in the end we have well determined centroid positions and
brightnesses for the template.

\subsection{Finding the shift}

For the subframes during the event, we want to be careful that we do not use
the source star or any of its nearby neighbors to compute the transformation
vector, since we {\em expect} a shift as a function of time.  As such, we
exclude the central $10\times 10$ pixels around the lensed star for this
calculation, but otherwise compute the transformation vector exactly as during
the unlensed period. Once the transformation matrix is computed we transform
the position and brightness of the source star to the template coordinate
system.

An estimate of the errors in the shift for each subframe was
determined by comparing stars similar (differing
by less than $20\%$) to the source star in brightness {\em as measured in that
frame} to their positions and brightnesses in the template catalog.  We
typically find $15-30$ comparison stars per subframe, and their distribution is
approximately Gaussian.  As an estimate of the uncertainties in position, we
take the dispersions around the known position of each comparison star.

In order to estimate the uncertainties in brightness, we used the magnitude
uncertainty given by DoPHOT, multiplied by a correctional coefficient which
takes the crowdedness into effect (Udalski et al. 1994a).  Typically, these
coefficients are $\sim 1.2$.

Finally, we compute a best fit to equation~(\ref{eq:lines}) and, where
applicable, equation~(\ref{eq:lcurve}) Note that we only use those data points
which occur {\em during} the microlensing event.  The unlensed frames are only
used in computing the baseline positions and brightnesses.  Uncertainties in
$\Delta x_{0}$ and $\Delta y_{0}$ are estimated by:

\begin{eqnarray}
\sigma_{\Delta x_{0}}\simeq \left( \frac{\partial \chi^{2}}{\partial \Delta
x_{0}} \right)^{-1} & \ ; & 
\sigma_{\Delta y_{0}}\simeq \left( \frac{\partial \chi^{2}}{\partial \Delta
y_{0}}  \right)^{-1}\ ,
\end{eqnarray}
evaluated around the best fit values.

The uncertainties in the light curve parameters are computed by determining a
covariance matrix for $u_{min}$, $t_{0}$ and $t_{max}$; the associated errors
are calculated in the standard way:
\begin{equation}
\sigma_{i}=\left[({\bf C}^{-1})_{ii}\right]^{1/2}\ ,
\end{equation}
where,
\begin{equation}
C_{ij}=\frac{\partial^2 \chi^2}{\partial \alpha_{i} \partial \alpha_{j}}\ ,
\end{equation}
and $\alpha_{i}$ is the set of light curve parameters.

\section{Results}
\label{sec:results}

Of the 19 candidate microlensing events reported by the OGLE collaboration,
subframes were reduced for all but OGLE $\# 13$, which was not part of the
normal search field, but rather was a followup of a MACHO event toward the
galactic bulge (Szyma\'nski et al. 1994).  OGLE $\# 2$ was on the boundary
between the BWC and BW5 fields, and hence, we have two sets of subframes which
are analyzed independently from one another.

Of the remaining events, OGLE $\#$'s $4$, $16$, and $19$ were not consistently
detected using DoPHOT.  This bears discussion.  Since we did not use a template
image to determine positions of stars, each frame was run independently, and
thus dim source stars were not always detected.  These stars had I magnitudes
of $\simeq 19.3, 18.5, \& 19.6$, respectively.  Likewise, the lack of a fixed
template allowed the centroid to ``move'' due to local effects.x  One of the
results of this is that, in general, we effectively had worse photometry than
in the official OGLE catalog.

We summarize the results of the $15$ observed events, including the double set
of OGLE $\# 2$ in table~\ref{tab:events}.  Throughout, all units of position
are given in units of pixel lengths.  Of the $15$ events, $7$ showed shifts of
$\geq 0.5$ pixels, not including OGLE $\# 12$, which will be discussed shortly.
Some, such as OGLE $\# 6$ had such a large and well-defined shift, that is
straightforward to pick out the slope even by eye.  We also show the plots of
the observed astrometric shift (in pixels) and the light curves for the events
in Figures 1 and 2, respectively.  It should further be noted that the
microlensing light curves that we determined were generally quite similar to
those computed by the OGLE collaboration.  On primary reason for difference in
our parameters may have been the use of a moving PSF.  Since the integrated
brightness is essentially a convolution of PSFs, it should be unsurprising that
allowing a moving PSF will also allow for variation in brightness.

Many of the events have interesting, occasionally unexpected, behavior.
First, there were $3$ events which had obvious outliers: OGLE $\#$'s $1$, $3$,
and $14$.  In each case, there were points which were several $\sigma$ from
the best fit of equation~(\ref{eq:lines}).  Moreover, photometrically, they
clearly differed from the other points in the light curve.  We removed the
outliers only where there was a poor fit both astrometrically and
photometrically.  For the sake of completeness, we have plotted the outliers as
open triangles in both Figures 1 and 2.  An inspection of the frames containing
the outliers reveal that they are primarily caused by a bad PSF, and CCD
defects.  

OGLE \# 2 was measured twice, and since the shift was significant, (and indeed,
consistent to within $1 \sigma $ errorbars in both the $x$ and $y$ directions),
this serves as an affirmation that we are, in fact, detecting a shift.
Moreover, it suggests that our estimates on our uncertainties were not
unreasonable.

OGLE $\# 5$ had a significant shift of $\sim 0.7$ pixels.  This is quite
encouraging as this event had been previously considered to be blended based on
photometric information (Alard 1997).

OGLE $\# 7$ is a binary event, which the OGLE collaboration determined to be
blended.  They were able to perform a best fit to the binary lens microlensing
parameters (Udalski et al. 1994c) using an observation of the event at the
caustic crossing.  We were unable to observe this point, and hence our
photometry would be insufficient to uniquely determine the blending of the
event.  We did, however, detect a small astrometric shift of $\simeq 0.16$
pixels.

Near OGLE $\# 12$, we found a nearby ($\sim 3$ pixels) dim star which was only
occasionally observed.  When it was not observed, the best fit of the source
star was shifted toward the dim star.  Thus, we find essentially two
populations of images, one in which the companion is observed ($12a$), and one
in which the companion is not observed ($12b$).  Visual inspection of the
frames reveals a CCD defect near the source image.  Since the CCD is not
aligned exactly the same with respect to the star field from night to night,
some observations have this defect overlapping the source (in which case, the
source is not observed), others have it nearby (in which case, the
``companion'' star is not observed), and others have it far away (in which case
both are observed).

The surprising thing about these two populations is that when the companion is
unobserved, and hence, when the event is {\em more} severely blended, we find
that the baseline brightness of the source is $\sim 10\%$ dimmer than when the
companion is observed.  We believe that this is due to the fact that the
centroid of the best fit PSF is shifted off the PSF of the source star, and
hence, a non-negligible fraction of the light is being thrown out by DoPHOT.

The positions for both of the cases are given in coordinates with the weighted
average center at the origin.  Taking the two populations separately, we find
that the center of $12a$ is $(0.35\ ,\ 0.29)$, and the center of $12b$ is at
$(-0.37\ ,\ -0.29)$.  Given that, the actual shifts for the two cases
respectively are $(-0.11\pm 0.09\ ,\ -0.15\pm 0.09)$ and $(0.67\pm 0.09 \ , \
0.82\pm 0.09)$.  Thus, 12a has almost no shift.  This, coupled with the fact
that we have subtracted out a nearby star, suggests that the remaining
measurement is essentially unblended.

Note that the values of $\Delta x_{0}$ and $\Delta y_{0}$ in
table~\ref{tab:events} are given in the same coordinate system, and hence we
expect the 12a and 12b to have identical values of both, as $\Delta x_{0}$ and
$\Delta y_{0}$ represent the physical position of the source star.  Indeed, in
the $x$ coordinate, the two agree quite well.  However, in the $y$ coordinate,
the two populations give inconsistent values to several $\sigma$.  This is
illustrated in Figure 1c, in which we would expect that the shift curves of the
two observations will intersect at $(A_{obs}-1)/A_{obs}=1$ (infinite
magnification), since observed position ought to be the same, whether or not
the nearby companion is also observed.

This event is extremely interesting in that it points out the importance of
accurate centroiding in photometric measurements.  Since the PSF centers were
fixed in previous studies, this has not been an issue.  However, if this
technique is to be more widely employed, the local environment, and the
presence or absence of nearby stars must be carefully monitored.

Our light curves for OGLE $\#'s 14$ \& $18$ are each extremely sparsely
populated and of very poor quality.  This is primarily due to the dimness
(I=$19, 18.7$, respectively) of the stars.  Though little
confidence should be placed in our computed values for the microlensing
parameters, we have included the curves for the sake of completeness.

\section{Discussion and Conclusions}
\label{sec:discussion}

The primary result of this analysis seems to be a positive one.  Astrometric
shifts were found convincingly in over half the candidate microlensing events,
some, such as OGLE $\# 6$, unambiguously detectable by eye.  This confirms the
belief that most events are significantly blended and that the majority of the
events will show some measurable shift.  Moreover, though we performed analysis
on a particular dataset (OGLE-1), using an off the shelf image processing
software package (DoPHOT), we suggest that the presence of an astrometric shift
is a generic one, and that all projects toward the Galactic bulge, the LMC, or
the SMC will show a significant fraction of events with large shifts.  Further
analysis could be used to provide an estimate of a statistical correction to
the optical depth of crowded fields to microlensing.

However, those events which show a shift may also be analyzed using followup
analysis on a high resolution ground-based instrument.  The detection of a
shift gives the position of the source star in local coordinates with very high
accuracy, and hence it becomes a straightforward matter to identify the correct
source star in a high resolution image, and from that determine the value of
$f$, and hence the other microlensing parameters, and the correction to the
optical depth.  Moreover, the very existence of a shift suggest which events
are expected to be blended.

However, in doing this, we encountered several difficulties.  First, we note
that the local effects (eg nearby stars of comparable brightness) pose great
difficulties in consistently measuring the position and brightness of the
lensed star.  Though this effect was pronounced in OGLE $\# 12$, it may have
played a smaller role in other events and thus positional measurements are
somewhat problematical.  

Additionally, both local effects, and a dim source star can cause the
star to be observed only intermittently.  As a result, our uncertainties in
$\Delta x_{0}$ and $\Delta y_{0}$ as well as the microlensing parameters are
not measured with as high a precision as if the star had been detected on each subframe.

Moreover, photon statistics, crowding, and variable seeing, and so on, caused
some difficulty in photometric measurements.  Indeed, this is far from the
ideal if one wishes to get optimum photometry.  It should be noted, though,
that the photometry does not artificially blend additional stars which were not
blended in the initial OGLE measurements.  If this were the case, we would
measure a smaller amplification than in the OGLE catalog, which we do not.

One possible (and encouraging) way around these difficulties in measurement is
frame subtraction (Alard \& Lupton 1997).  This method will ultimately provide
a direct unshifted position of the source and unblended light in every frame.

In conclusion, we have shown that most events toward the galactic bulge do,
in fact, have a significant astrometric shift.  Moreover, we can conclude a
high degree of blending.  It is hoped that followup observations of this and
similar analyses can help to provide greater accuracy in the measurement of
microlensing optical depths.

\acknowledgements{ The authors would like to gratefully acknowledge Bohdan
Paczy\'nski for invaluable discussions, and the OGLE team, in particular
Andrzej Udalski and Micha\l ~Szyma\'nski for allowing us the use of their data.
This research was supported by NSF grant AST-9530478, and DMG was supported by
an NSF graduate research fellowship.}

\newpage

{\footnotesize

\begin{planotable}{rrrrrrrrr}
\tablewidth{43pc}
\tablecaption{A summary of the astrometric shifts and
microlensing parameters of the observed OGLE events.  Column 1 shows the event
$\#$ used in the OGLE literature.  Column 3 shows the number of frames in which
the star is observed to be lensed.  Columns 4 and 5 show the computed
astrometric shift (equations~[\ref{eq:lines}]) in pixels, and the remaining
columns give the computed microlensing parameters.  Column 8 is the estimated
date of maximum amplification in (J.D.-2448000).}

\tablehead{\colhead{$\#$} & \colhead{Star
$\#$} & \colhead{obs.} & \colhead{$\Delta x_{0}$} & \colhead{$\Delta y_{0}$} &
\colhead{$A_{obs}^{max}$} & \colhead{$t_{0}$ (days)} &
\colhead{$t_{max}$ (days)} & \colhead{$\chi^{2}_{min}$}} \startdata

$1$\footnotemark[1]
& BW7 117281 & 38 & $-0.58\pm 0.04$ & $-0.11\pm 0.04$ &
$2.60\pm 0.10$ & $21.3\pm 1.2$ & $1154.6\pm 0.8$ & $17.6$ \nl

$2a$ & BW5 178651 & 26 & $-0.39\pm 0.04$ & $-0.53\pm 0.04$ &
$8.51\pm 0.37$ & $52.5\pm 1.8$ & $803.3\pm 0.6$ & $21.9$ \nl

$2b$ & BWC 10648 & 27 & $-0.49\pm 0.04$ & $-0.49\pm 0.04$ &
$7.53\pm 0.35$ & $48.2\pm 1.4$ & $805.0\pm 0.5$ & $34.7$ \nl

$3$\footnotemark[1]
   & BW3 161225 & 44 & $0.25\pm 0.07$ & $-0.18\pm 0.07$ &
$1.30\pm 0.02$ & $14.4\pm 1.5$ & $833.7\pm 1.1$ & $92.8$ \nl

$5$ & BWC 120698 & 20 & $0.64\pm 0.02$ & $-0.34\pm 0.02$ &
$13.58\pm 0.16$ & $12.7\pm 0.16$ & $824.4\pm 0.2$ & $85.1$ \nl

$6$ & MM5-B 128727 & 30 & $0.86\pm 0.06$ & $0.09\pm 0.05$ &
$7.78\pm 0.3$ & $9.3\pm 0.2$ & $818.7\pm 0.1$ & $53.6$ \nl

$7$\footnotemark[2]
  & BW8 198503 & 29 & $-0.04\pm 0.01$ & $0.16\pm 0.01$ &
 & & & \nl

$8$ & BW9 138910 & 13 & $-0.83\pm 0.09$ & $0.93\pm 0.12$ &
$1.56\pm 0.05$ & $37.0\pm 5.9$ & $1217.7\pm 5.1$ & $2.4$ \nl

$9$ & MM7-A 86776 & 11 & $-0.30\pm 0.14$ & $0.76\pm 0.15$ & 
$1.75\pm 0.14$ & $16.4\pm 4.9$ & $815.0\pm 1.5$ & $6.5$ \nl

$10$ & BW3 161220 & 44 & $-0.08\pm 0.08$ & $-0.05\pm 0.09$ &
$1.11 \pm 0.01$ & $103.0\pm 21.4$ & $840.1\pm 23.0$ & $113.1$ \nl

$11$ & BW6 167045 & 33 & $-0.08\pm 0.10$ & $-0.46\pm 0.15$ &
$1.28\pm 0.03$ & $10.4\pm 2.0$ & $1536.8\pm 1.9$ & $16.4$ \nl

$12a$\footnotemark[3]
   & BW5 83758 & 27 & $0.23\pm 0.09$ & $0.14\pm 0.09$ &
$1.94\pm 0.09$ & $17.8\pm 1.4$ & $1583.1\pm 1.0$ & $30.9$ \nl

$12b$\footnotemark[3]
   & BW5 83758 & 34 & $0.30\pm 0.09$ & $0.55\pm 0.09$ &
$2.13\pm 0.10$ & $22.1\pm 1.4$ & $1581.6\pm 0.9$ & $74.0$ \nl

$14$\footnotemark[1]
    & MM1-A 123474 & 5 & $-0.5\pm 0.15$ & $-0.43\pm 0.15$ &
$2.30\pm 0.36$ & $25.0\pm 10.7$ & $1815.9\pm 7.8$ & $8.4$ \nl

$15$ & BW3 142477 & 19 & $0.46\pm 0.05$ & $-0.64\pm 0.05$ &
$4.81\pm 1.68$ & $16.9\pm 0.9$ & $1854.2\pm 0.8$ & $16.4$ \nl

$17$ & BW10 176006 & 30 & $0.00\pm 0.10$ & $-0.05\pm 0.07$ &
$1.57\pm 0.15$ & $139.9\pm 52.5$ & $1975.7\pm 55.9$ & $41.2$ \nl

$18$ & BW1 67895 & 6 & $-0.07\pm 0.25$ & $3.35\pm 0.35$ &
$2.09\pm 6.6$ & $21.2\pm 21.7$ & $1914.5\pm 43.3$ & $0.43$ \nl
\enddata

\label{tab:events}
\end{planotable}

$^{1}$ OGLE $\#$'s $1$m $3$ \& $14$ each have $1$, $3$, and $3$ 
obvious outliers, respectively. In both cases, the outliers are both
photometrically and astrometrically distinct from the rest of the population.

$^{2}$ OGLE $\# 7$ is a binary event.

$^{3}$ Measurements of OGLE $\# 12$ are highly dependent upon detection of a
	nearby faint star.  The interpretation of ``events'' $12a$ and $12b$
	are discussed in the text

\clearpage

%Put in figures here...

\newcounter{dumb}
\newcounter{temp}
\setcounter{temp}{\value{figure}}
\setcounter{figure}{0}
\setcounter{dumb}{1}
\renewcommand{\thedumb}{\arabic{dumb}}
\renewcommand{\thefigure}{\thedumb \alph{figure}}

\begin{figure}
\centerline{\psfig{figure=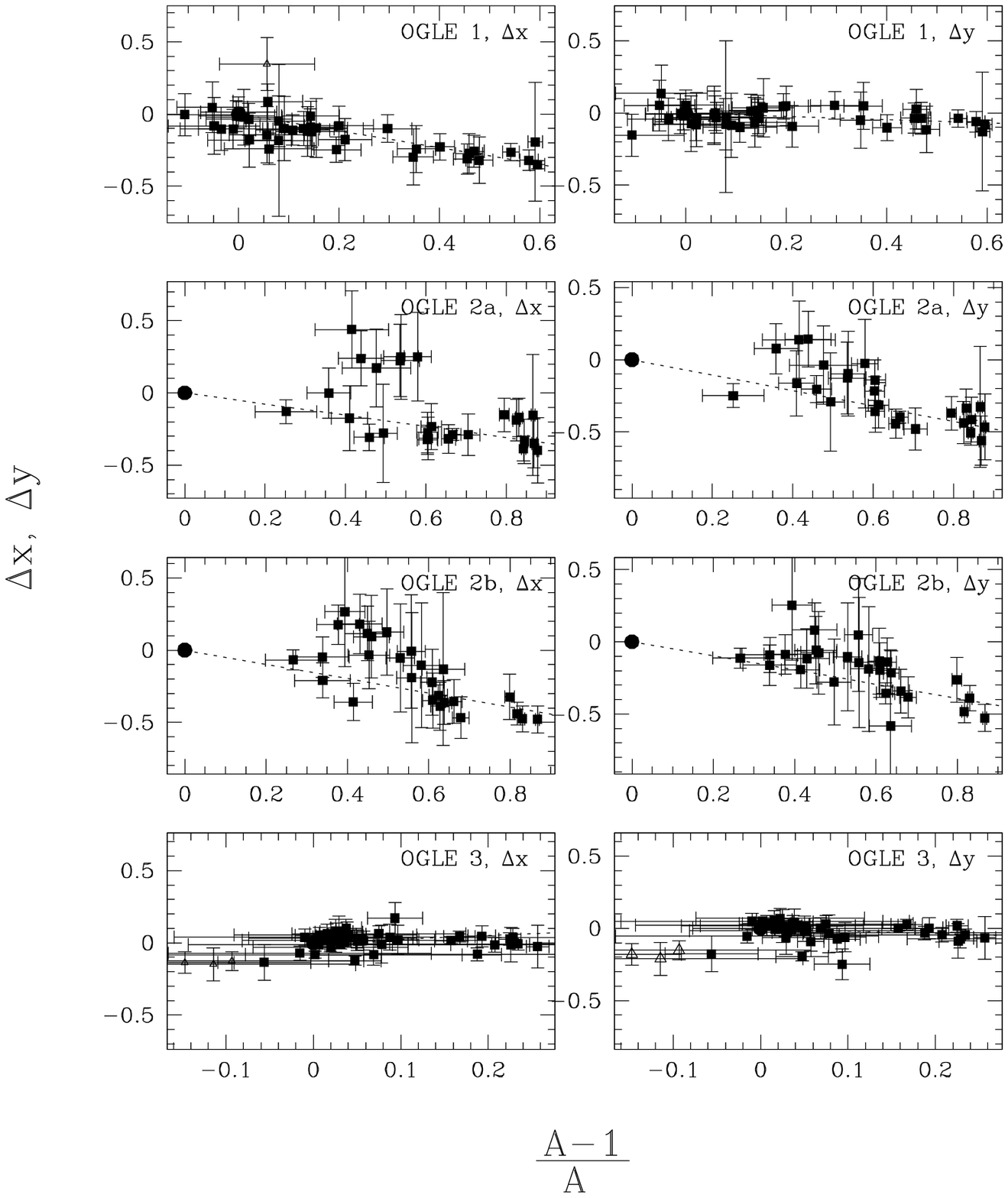,height=8in,angle=0}}
\caption{The astrometric shifts in OGLE $\#'s 1-3$.  The dotted line shows the
best fit to equations (4), yielding the values of $\Delta x_{0}$ and $\Delta
y_{0}$ in Table 1.   The solid circle shows the position and brightness of the
source during the unlensed state.  }
\end{figure}

\begin{figure}
\centerline{\psfig{figure=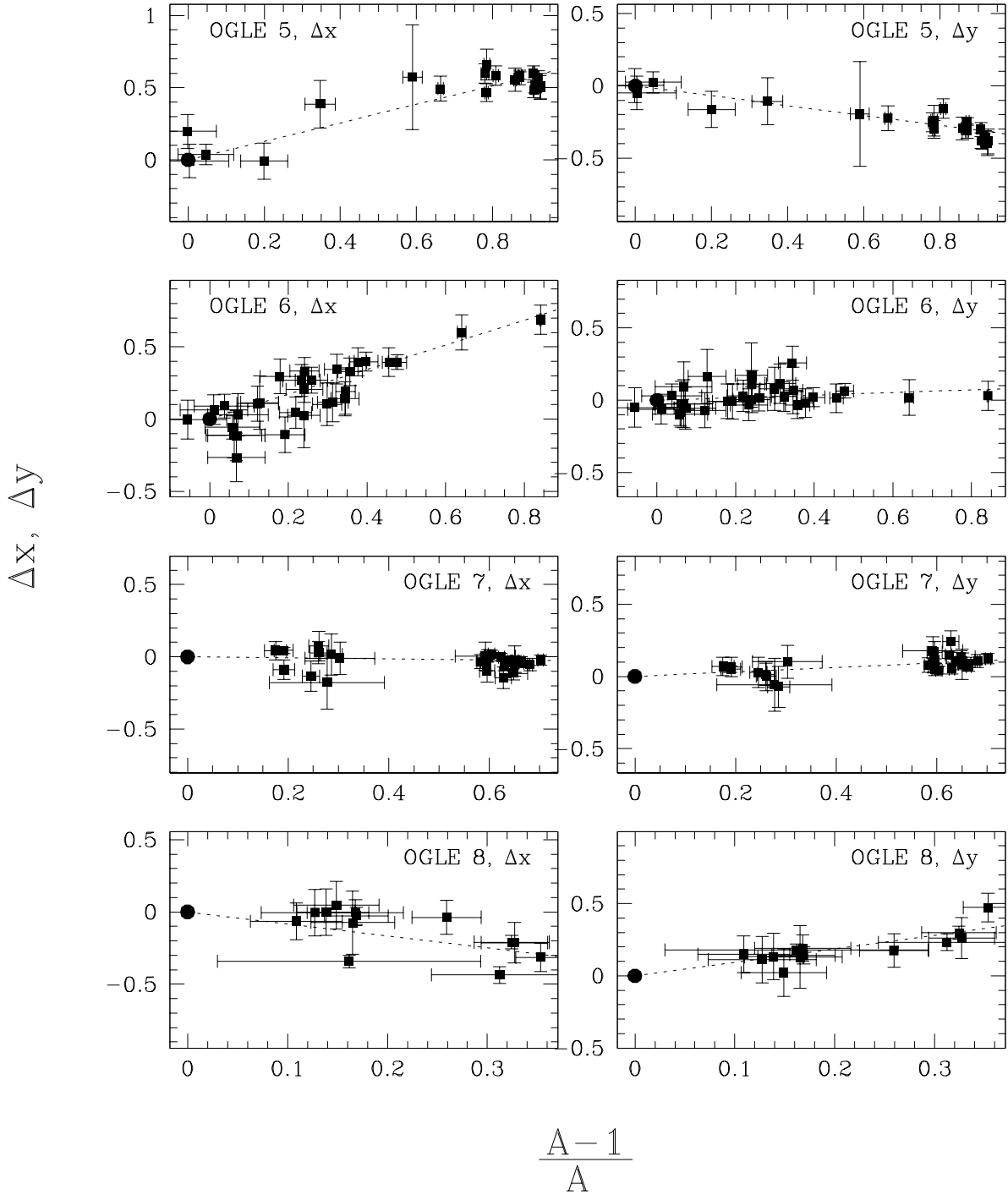,height=8in,angle=0}}
\caption{As above, the astrometric shifts in OGLE $\#'s 5-8$.}
\end{figure}

\begin{figure}
\centerline{\psfig{figure=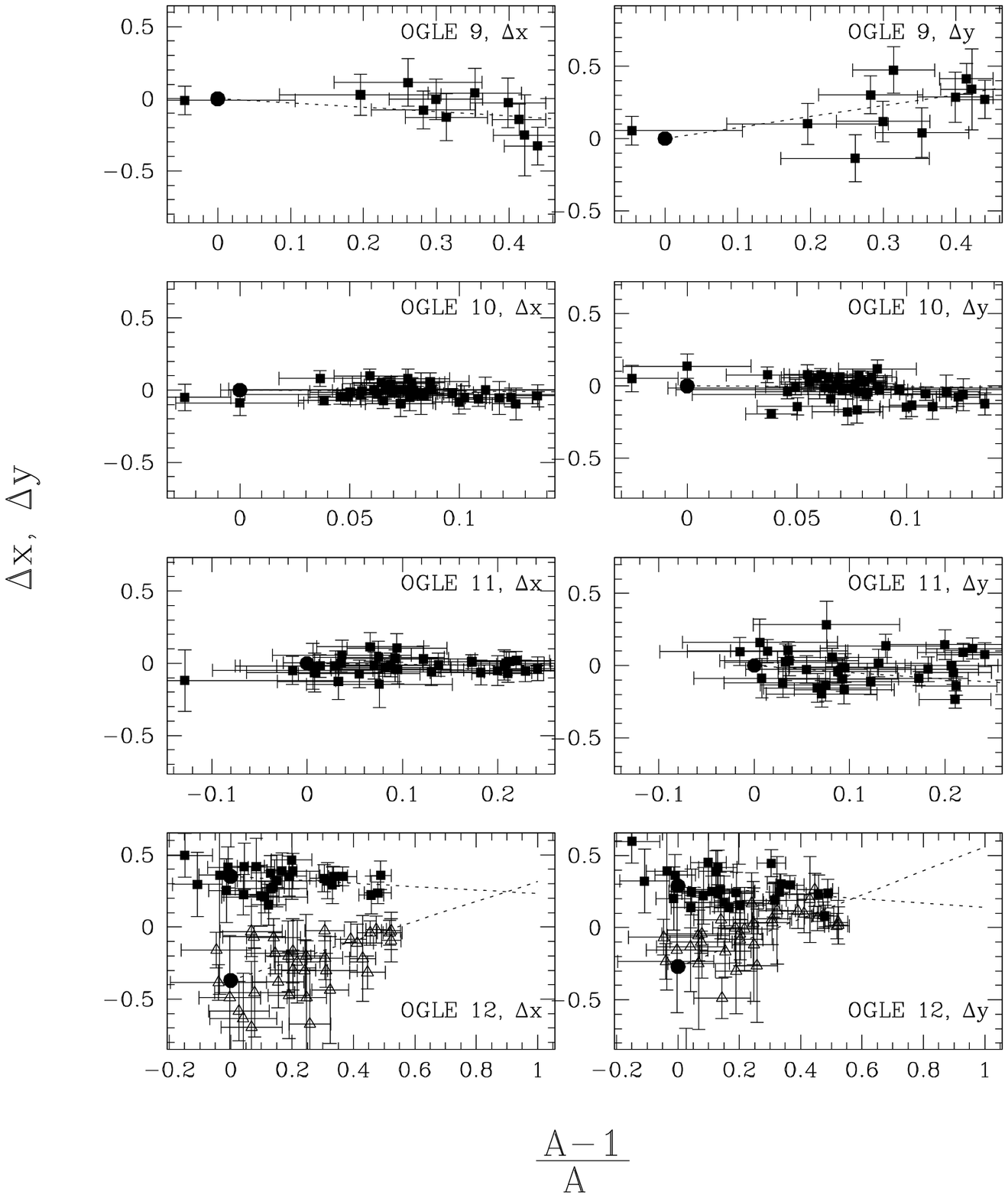,height=8in,angle=0}}
\caption{As above, the astrometric shift in OGLE $\#'s 9-12$.}
\end{figure}

\begin{figure}
\centerline{\psfig{figure=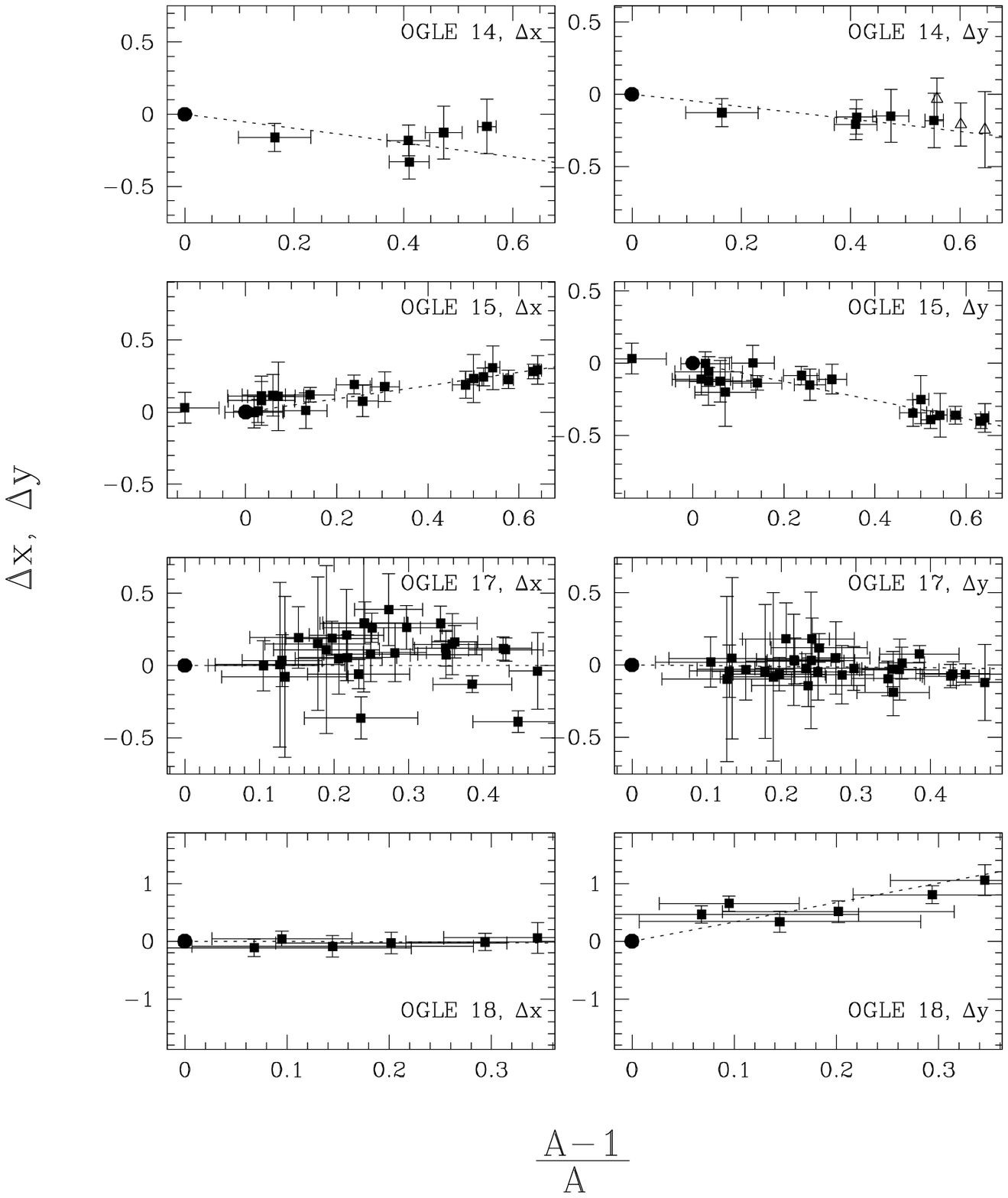,height=8in,angle=0}}
\caption{As above, the astrometric shift in OGLE $\#'s 14-18$.}
\end{figure}

\setcounter{figure}{\value{temp}}
\renewcommand{\thefigure}{\arabic{figure}}

\setcounter{temp}{\value{figure}}
\setcounter{figure}{0}
\setcounter{dumb}{2}
\renewcommand{\thedumb}{\arabic{dumb}}
\renewcommand{\thefigure}{\thedumb \alph{figure}}

\begin{figure}
\centerline{\psfig{figure=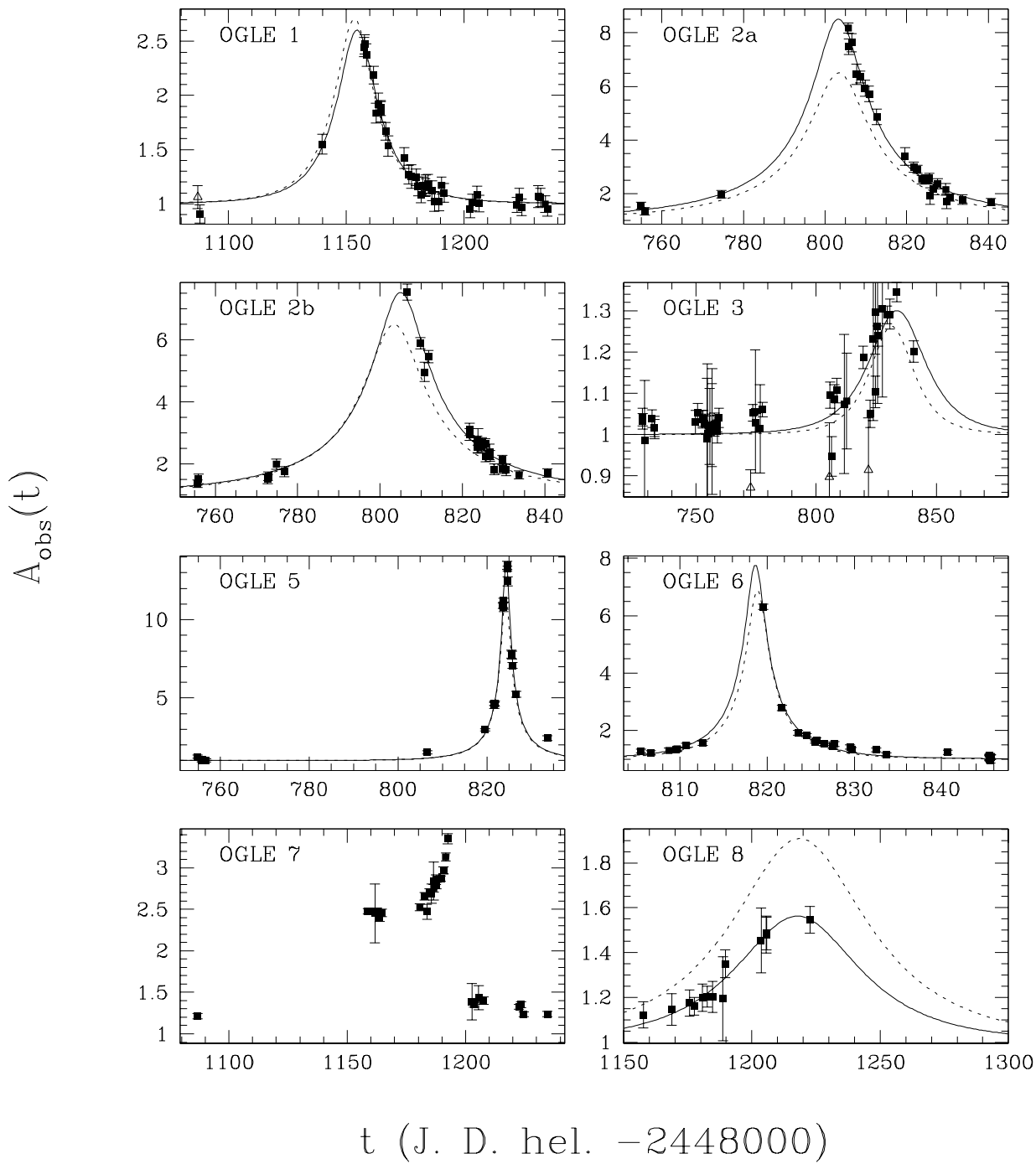,height=8in,angle=0}}
\caption{The light curves for OGLE $\#'s 1-8$.  The solid line shows the newly
calculated parameters, while the dotted line shows the OGLE parameters.  The
solid squares are our measurements of the magnification.  OGLE $7$ is a binary,
for which we have not calculated microlensing parameters.}
\end{figure}

\begin{figure}
\centerline{\psfig{figure=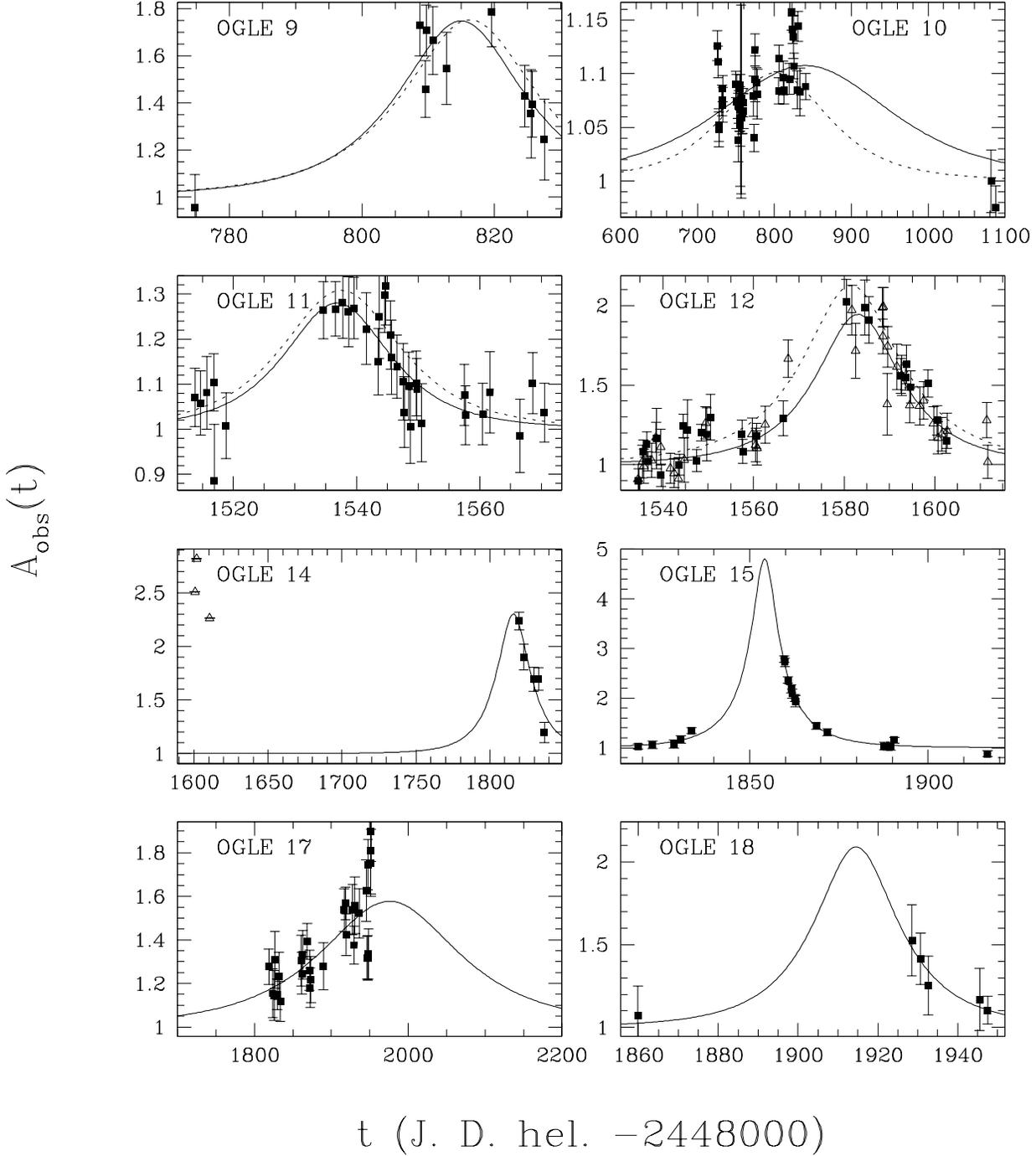,height=8in,angle=0}}
\caption{As above, the light curve for OGLE $\#'s 9-18$.  Note, in OGLE 12, the
solid squares and solid line correspond to OGLE 12a, the ``unblended''
realization discussed in the text.  The dotted line and open triangles
correspond to the case in which the nearby companion was not detected.  For
OGLE $\#'s 14-18$, the OGLE group had not previously published a light curve.}
\end{figure}

\setcounter{figure}{\value{temp}}
\renewcommand{\thefigure}{\arabic{figure}}

\end{document}